%% file: main.tex
\documentclass[sigconf]{acmart}
\renewcommand\footnotetextcopyrightpermission[1]{} 
\usepackage{fancyhdr}
\usepackage[normalem]{ulem}
\usepackage{xspace}
\usepackage{graphicx}
\usepackage{xcolor}
\usepackage{bbold}

\usepackage{amsmath,amsthm,mathtools}
\usepackage{multirow}
\usepackage[caption=false]{subfig}
\usepackage{flushend}
\usepackage[utf8x]{inputenc}
\usepackage[ruled]{algorithm2e} 

\begin{document}
\title{Loom: Exploiting Weight and Activation Precisions to Accelerate Convolutional Neural Networks}

 \author{Sayeh Sharify, Alberto Delmas Lascorz, Kevin Siu, Patrick Judd,  Andreas Moshovos}
 \affiliation{%
   \institution{University of Toronto}
 }
 \email{{sayeh, delmasl1, siukevi4, juddpatr, moshovos}@ece.utoronto.ca}




\newcommand{\STRD}{\textit{DStripes}\xspace}
\newcommand{\STR}{\textit{Stripes}\xspace}
\newcommand{\BSDL}{\textit{Loom}\xspace}
\newcommand{\BSD}{\textit{LM}\xspace}
\newcommand{\DaDN}{\textit{DaDN}\xspace}
\newcommand{\BASE}{\textit{DPNN}\xspace}
\newcommand{\FCLs}{FCLs\xspace}
\newcommand{\CVLs}{CVLs\xspace}
\newcommand{\DSTRTITLE}{{DPRS}\xspace}
\newcommand{\meanspeedupFCstr}{$1.0\times$\xspace}
\newcommand{\meanspeedupCVstr}{$2.44\times$\xspace}
\newcommand{\meanspeedupALLstr}{$2.38\times$\xspace} 
\newcommand{\meanspeedupFCone}{$1.74\times$\xspace}
\newcommand{\meanspeedupCVone}{$3.25\times$\xspace}
\newcommand{\meanspeedupALLone}{$3.19\times$\xspace} 
\newcommand{\meanspeedupFC}{$1.75\times$\xspace}
\newcommand{\meanspeedupCV}{$3.10\times$\xspace}
\newcommand{\meanspeedupALL}{$3.05\times$\xspace} 
\newcommand{\meanspeedupFCfour}{$1.75\times$\xspace}
\newcommand{\meanspeedupCVfour}{$2.78\times$\xspace}
\newcommand{\meanspeedupALLfour}{$2.74\times$\xspace}
\newcommand{\meanefficiencyFCstr}{$0.86\times$\xspace}
\newcommand{\meanefficiencyCVstr}{$2.11\times$\xspace}
\newcommand{\meanefficiencyALLstr}{$2.06\times$\xspace}
\newcommand{\meanefficiencyFCone}{$1.41\times$\xspace}
\newcommand{\meanefficiencyCVone}{$2.63\times$\xspace}
\newcommand{\meanefficiencyALLone}{$2.59\times$\xspace} 
\newcommand{\meanefficiencyFC}{$1.65\times$\xspace}
\newcommand{\meanefficiencyCV}{$2.92\times$\xspace}
\newcommand{\meanefficiencyALL}{$2.87\times$\xspace} 
\newcommand{\meanefficiencyFCfour}{$1.84\times$\xspace}
\newcommand{\meanefficiencyCVfour}{$2.92\times$\xspace}
\newcommand{\meanefficiencyALLfour}{$2.89\times$\xspace} 
\newcommand{\meanspeedupFClossone}{$1.85\times$\xspace}
\newcommand{\meanspeedupCVlossone}{$3.63\times$\xspace}
\newcommand{\meanspeedupALLlossone}{$3.57\times$\xspace} 
\newcommand{\meanspeedupFCloss}{$1.85\times$\xspace}
\newcommand{\meanspeedupCVloss}{$3.45\times$\xspace}
\newcommand{\meanspeedupALLloss}{$3.40\times$\xspace} 
\newcommand{\meanspeedupFClossfour}{$1.86\times$\xspace}
\newcommand{\meanspeedupCVlossfour}{$3.11\times$\xspace}
\newcommand{\meanspeedupALLlossfour}{$3.07\times$\xspace} 
\newcommand{\meanefficiencyFClossone}{$1.49\times$\xspace}
\newcommand{\meanefficiencyCVlossone}{$2.93\times$\xspace}
\newcommand{\meanefficiencyALLlossone}{$2.87\times$\xspace}  
\newcommand{\meanefficiencyFCloss}{$1.75\times$\xspace}
\newcommand{\meanefficiencyCVloss}{$3.25\times$\xspace}
\newcommand{\meanefficiencyALLloss}{$3.19\times$\xspace}  
\newcommand{\meanefficiencyFClossfour}{$1.95\times$\xspace}
\newcommand{\meanefficiencyCVlossfour}{$3.26\times$\xspace}
\newcommand{\meanefficiencyALLlossfour}{$3.18\times$\xspace}  

\newcommand{\fixme}[1]{{\textcolor{magenta} {#1}} }

\begin{abstract}
\BSDL (\BSD), a hardware inference accelerator for Convolutional Neural Networks (CNNs) is presented. In \BSD every bit of data precision that can be saved translates to proportional performance gains. Specifically, for convolutional layers \BSD's execution time scales inversely proportionally with the precisions of both weights and activations. For fully-connected layers \BSD's performance scales inversely proportionally with the precision of the weights. 
\BSD targets area- and bandwidth-constrained System-on-a-Chip designs such as those found on mobile devices that cannot afford the multi-megabyte buffers that would be needed to store each layer on-chip. Accordingly, given a data bandwidth budget, \BSD boosts energy efficiency and performance over an equivalent bit-parallel accelerator. 
For both weights and activations \BSD can exploit profile-derived per layer precisions. However, at runtime \BSD further trims activation precisions at a much smaller than a layer granularity. Moreover, it can naturally exploit weight precision variability at a smaller granularity than a layer. On average, across several image classification CNNs and for a configuration that can perform the equivalent of 128 $16b\times16b$ multiply-accumulate operations per cycle \BSD outperforms a state-of-the-art bit-parallel accelerator~\cite{DaDiannao} by $4.38\times$ without any loss in accuracy while being $3.54\times$ more energy efficient. \BSD can trade-off accuracy for additional improvements in execution performance and energy efficiency and compares favorably to an accelerator that targeted only activation precisions. We also study 2- and 4-bit L\BSD variants and find the the 2-bit per cycle variant is the most energy efficient. 
\end{abstract}

\maketitle

\input{samplebody-conf}

\end{document}

%% file: samplebody-conf.tex
\section{Introduction}

\label{sec:intro}

Deep neural networks (DNNs) have become the state-of-the-art technique in many recognition tasks such as object~\cite{RCNN13} and speech recognition~\cite{deep-speech}. Given their many applications and high computation and memory demands, DNNs are prime candidates for hardware acceleration. While a few different types of DNNs exist, 
Convolutional Neural Networks (CNNs) in particular dominate applications where the input is an image or video. Devices executing such CNNs will be required to perform mostly if not only inference. An example is computational photography where  machine learning has shown great promise in replacing classical algorithms~\cite{lukac2016computational}.
       
We present \BSDL (\BSD), a hardware accelerator for inference with CNNs targeting embedded systems where reducing the amount of data transfered per memory connection, be it an external or internal one, is paramount. Specifically, given a memory bandwidth budget \BSD's goal is to boost performance and energy efficiency compared to a state-of-the-art data-parallel accelerator. 
\hspace*{0.75mm} \BSD exploits the precision requirement variability of CNNs to reduce the memory footprint, increase bandwidth utilization, and to deliver performance which scales inversely proportional with precision for both convolutional (\CVLs) and fully-connected (\FCLs) layers. Ideally, compared to using a fixed precision of 16 bits, \BSD achieves a speedup of $\frac{256}{P_a\times P_w}$ and $\frac{16}{P_w}$ for \CVLs and \FCLs  where $P_{w}$ and $P_{a}$ are the precisions of weights and activations, respectively. \BSD also reduces the number of weight and activation bits read by $\frac{16-P_{w}}{16}$ and $\frac{16-P_{a}}{16}$.
To deliver these benefits \BSD processes both activations and weights bit-serially while compensating for the loss in computation bandwidth by exploiting parallelism. Judicious reuse of activations and weights enables \BSD to improve performance and energy efficiency over conventional bit-parallel designs without requiring a wider memory interface. For both weights and activations \BSD utilizes profile-derived per layer precisions. For activations, \BSD further trims their precision at a much finer granularity at runtime utilizing the approach of Lascorz \textit{et al.}~\cite{dynamicstripes}.
By exploiting precision \BSD delivers benefits for \textit{all} activations and weights regardless of whether they are ineffectual or not.

We evaluate \BSD on an SoC and compare against a bit-parallel fixed-precision accelerator (\BASE) over a set of image classification CNNs. 
 For a configuration that is sized to match the peak computation bandwidth of a bit-parallel accelerator that can perform at peak 128 $16b\times16b$ multiply-accumulate operations per cycle, on average \BSD yields a speedup of \meanspeedupCVone, \meanspeedupFCone, and \meanspeedupALLone over \BASE for the convolutional, fully-connected, and all layers, respectively. The energy efficiency of \BSD over \BASE is \meanefficiencyCVone, \meanefficiencyFCone and \meanefficiencyALLone for the aforementioned layers, respectively. \BSD enables trading off accuracy for additional improvements in performance and energy efficiency. For example, accepting a 1\% relative loss in accuracy, \BSD yields \meanspeedupALLlossone higher performance and \meanefficiencyALLlossone more energy efficiency than \BASE. We also perform a sensitivity study varying the equivalent peak compute bandwidth and the number of bits that \BSD processes per cycle. \BSD scales well up up to a configuration equivalent to 256 $16b\times16b$ multiply-accumulate operations per cycle and that a 2-bit per cycle design achieves the best energy efficiency albeit not the best performance.
 
The rest of this document is organized as follows: 
Section~\ref{sec:simple} illustrates the key concepts behind \BSD via an example. Section~\ref{sec:bsdArch} presents the \BASE and \BSDL architectures. The evaluation methodology and experimental results are presented in Section~\ref{sec:evaluation}. Section~\ref{sec:related} reviews related work, and Section~\ref{sec:conclusion} concludes.

\section{Loom: A Simplified Example}
\label{sec:simple}
This section explains how \BSD would process \CVLs and FCLs on an example using 2-bit activations and weights.

\begin{figure*}[!t]
\centering
\subfloat[Bit-Parallel Engine processing $2b \times 2b$ layer over two cycles]{
\centering
\includegraphics[width=0.28\textwidth]{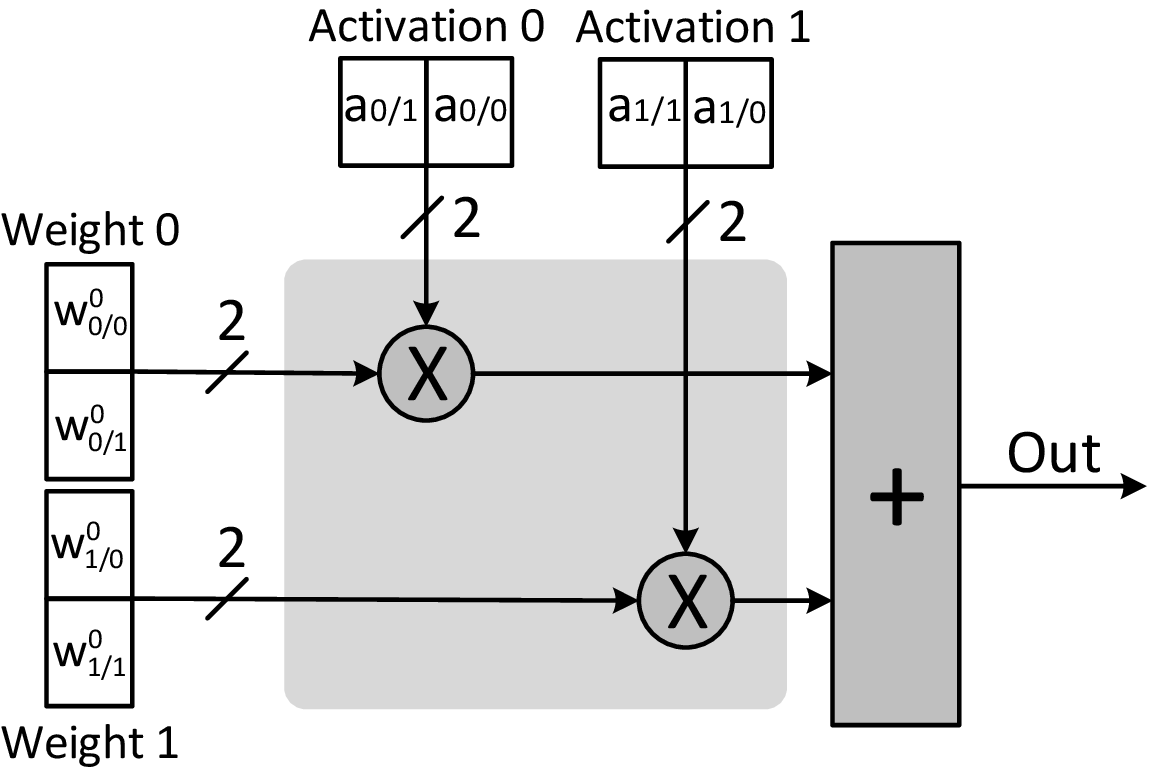}
\label{fig:bitSerial_FC_example_a}
}
\hspace{3pt}
\subfloat[Cycle 1: Load LSB of weights from filters 0 and 1 into the left WRs]{
\centering
\includegraphics[width=0.28\textwidth]{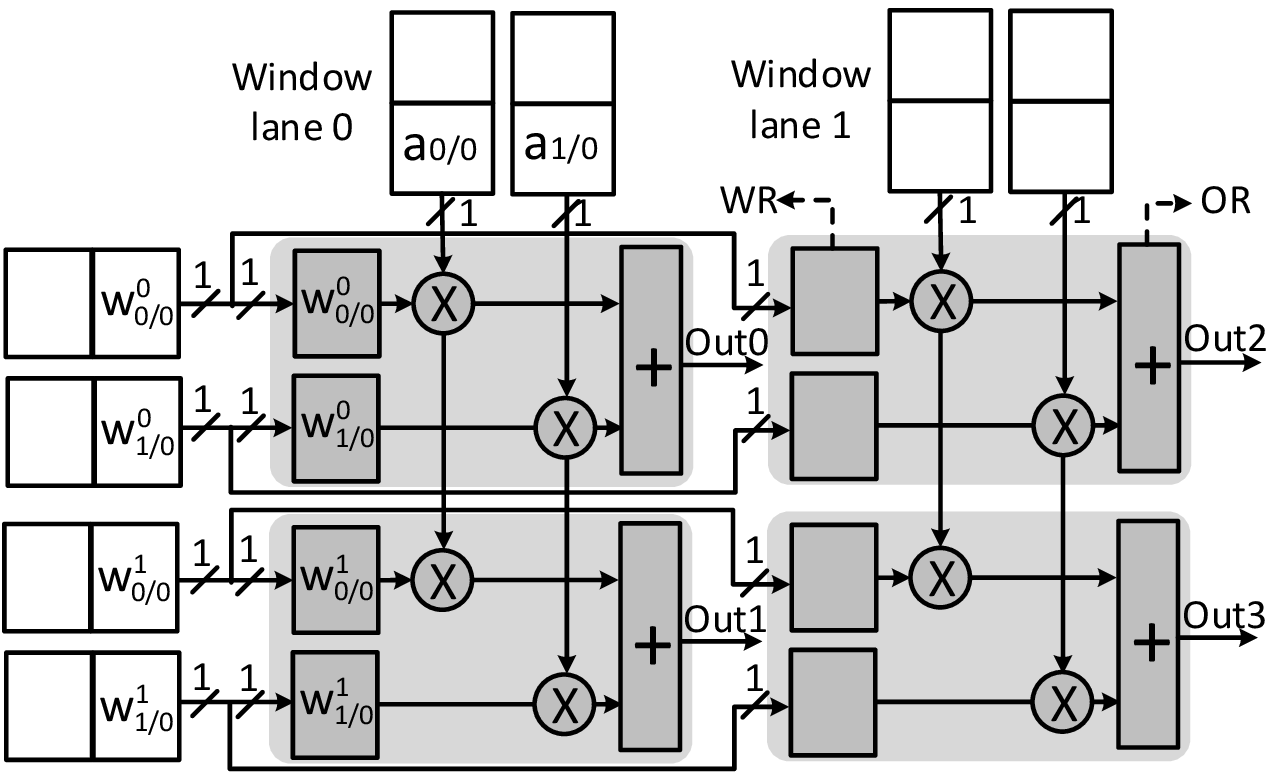}
\label{fig:bitSerial_FC_example_b}
}
\hspace{3pt}
\subfloat[Cycle 2: Load LSB of weights from filters 2 and 3 into the right WRs]{
\centering
\includegraphics[width=0.28\textwidth]{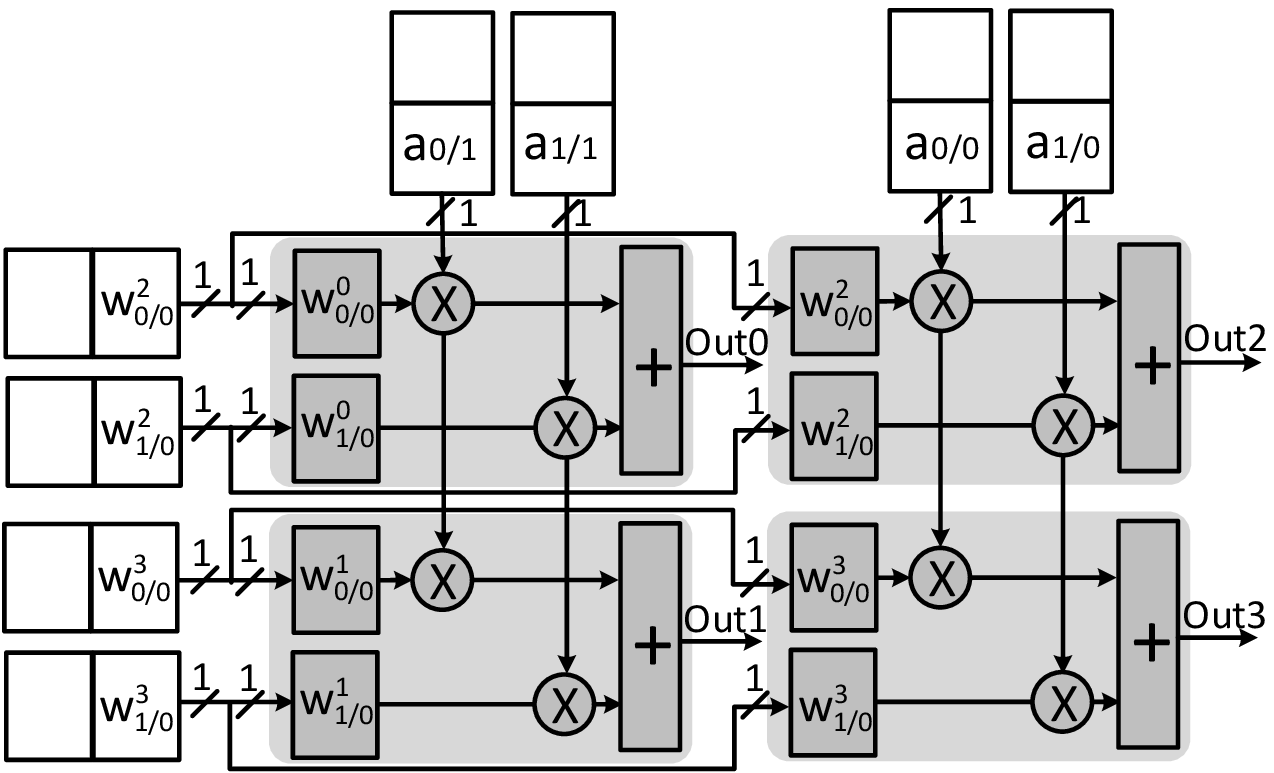}
\label{fig:bitSerial_FC_example_c}
}\\
\subfloat[Cycle 3: Load MSB of weights from filters 0 and 1 into the left WRs]{
\centering
\includegraphics[width=0.28\textwidth]{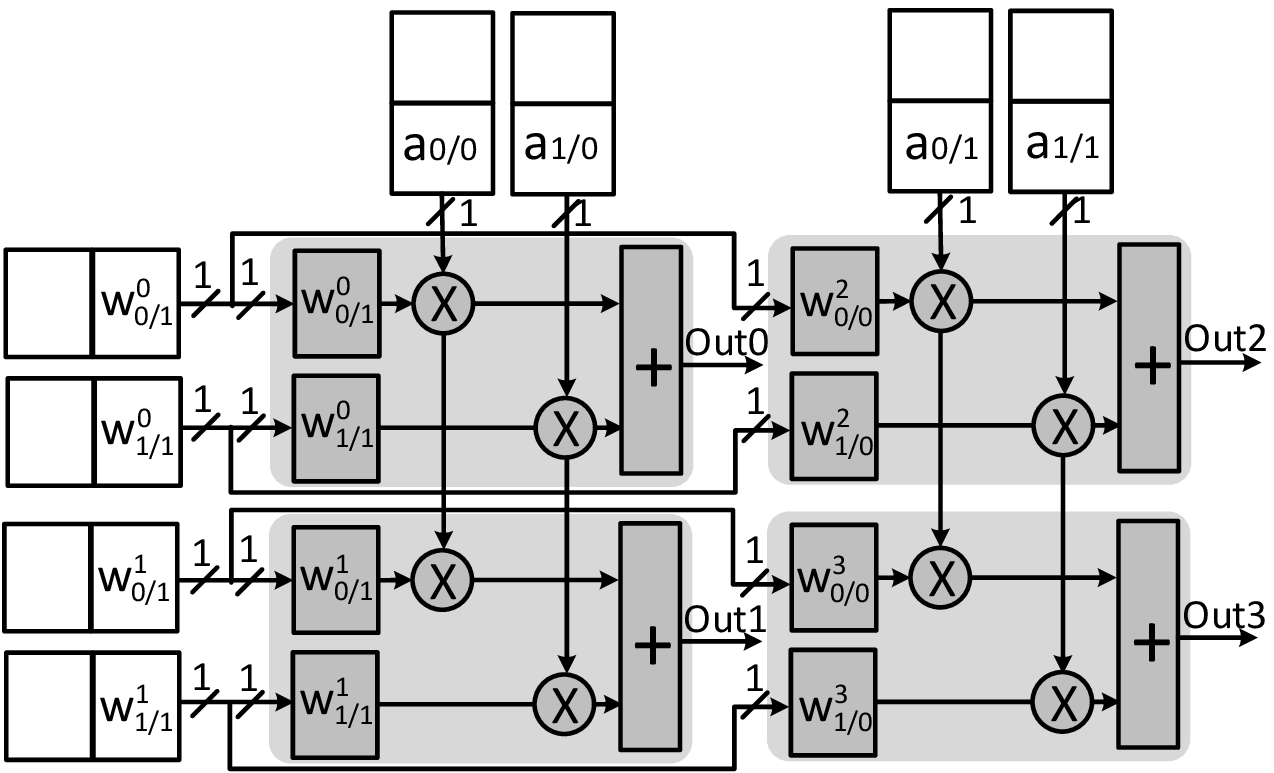}
\label{fig:bitSerial_FC_example_d}
}
\hspace{3pt}
\subfloat[Cycle 4: Load MSB of weights from filters 2 and 3 into the right WRs]{
\centering
\includegraphics[width=0.28\textwidth]{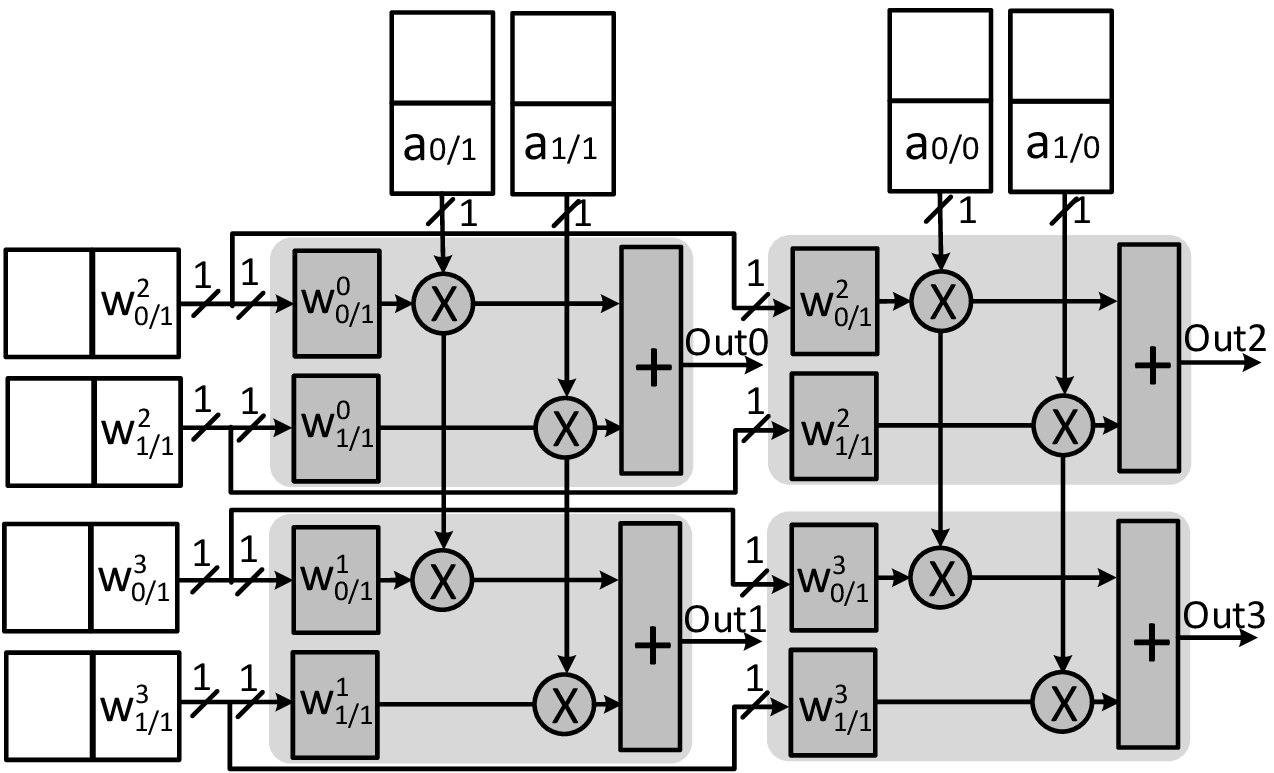}
\label{fig:bitSerial_FC_example_e}
}
\hspace{3pt}
\subfloat[Cycle 5: Multiply MSB of weights from filters 2 and 3 with MSB of $a_{0}$ and $a_{1}$]{
\centering
\includegraphics[width=0.28\textwidth]{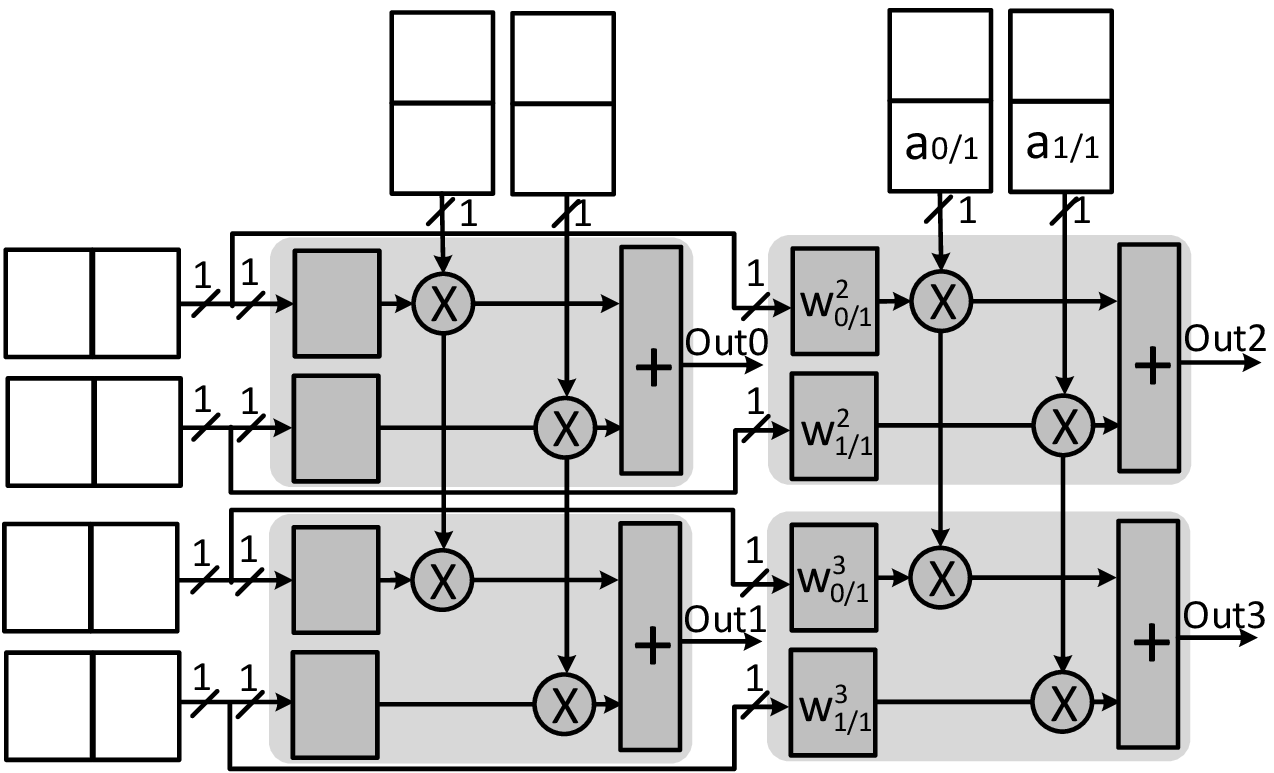}
\label{fig:bitSerial_FC_example_f}
}
\caption{Processing an example Fully-Connected Layer using \BSD's Approach.}
\label{fig:bitSerial_FC_example}
\end{figure*}

\noindent\textbf{Conventional Bit-Parallel Processing: }
Figure \ref{fig:bitSerial_FC_example_a} shows a bit-parallel processing engine which multiplies two input activations with two weights generating a single 2-bit output activation per cycle. The engine can process two new 2-bit weights and/or activations per cycle a  throughput of two $2b \times 2b$ products per cycle.

\noindent\textbf{\BSDL's Approach: }
Figure~\ref{fig:bitSerial_FC_example_b} shows an equivalent \BSD engine which matches the bit-parallel engine's throughput by producing 8 $1b \times 1b$ products every cycle. The engine comprises an $2\times 2$ array of bit-serial subunits (4 in  total). Each subunit accepts 2 bits of input activations and 2 bits of weights per cycle and performs 2 $1b\times 1b$ products. The subunits along the same column share the activation inputs while the subunits along the same row share their weight inputs. In total, this engine accepts 4 activation and 4 weight bits equaling the input bandwidth of the bit-parallel engine. Each subunit has two 1-bit Weight Registers (WRs), one 2-bit Output Register (OR) for accumulating its  products.

Figure~\ref{fig:bitSerial_FC_example_b} through Figure~\ref{fig:bitSerial_FC_example_f} show how \BSD would process an FCL. 
As Figure~\ref{fig:bitSerial_FC_example_b} shows, in \textbf{cycle 1}, the left column subunits receive the least significant bits (LSBs) $a_{0/0}$ and $a_{1/0}$ of activations $a_0$ and $a_1$, and $w^0_{0/0}$, $w^0_{1/0}$, $w^1_{0/0}$, and $w^1_{1/0}$, the LSBs of four weights from filters 0 and 1. Each of these two subunits calculates two $1b\times 1b$ products (the product and accumulation would take place in the subsequent cycle adding one more pipeline stage, a detail the example omits for clarity) and stores their sum into its OR. 
  In Figure~\ref{fig:bitSerial_FC_example_c} and \textbf{cycle 2}, the left column subunits now multiply the same weight bits with the most significant bits (MSBs) $a_{0/1}$ and $a_{1/1}$ of activations $a_0$ and $a_1$ respectively accumulate these into their ORs. In parallel, the two right column subunits load $a_{0/0}$ and $a_{1/0}$, the LSBs of the input activations $a_0$ and $a_1$, and multiply them by the LSBs of weights $w^2_{0/0}$, $w^2_{1/0}$, $w^3_{0/0}$, and $w^3_{1/0}$ from filters 2 and 3. In \textbf{cycle 3}, the left column subunits now load and multiply the LSBs $a_{0/0}$ and $a_{1/0}$ with the MSBs $w^0_{0/1}$, $w^0_{1/1}$, $w^1_{0/1}$, and $w^1_{1/1}$ of the four weights from filters 0 and 1. In parallel, the right subunits reuse their WR-held weights $w^2_{0/0}$, $w^2_{1/0}$, $w^3_{0/0}$, and $w^3_{1/0}$ and multiply them by the most significant bits $a_{0/1}$ and $a_{1/1}$ of activations $a_0$ and $a_1$ (Figure~\ref{fig:bitSerial_FC_example_d}).  In \textbf{cycle 4} and Figure~\ref{fig:bitSerial_FC_example_e}, the left column subunits multiply their WR-held weights and $a_{0/1}$ and $a_{1/1}$ the MSBs of activations $a_0$ and $a_1$ and finish the calculation of output activations $o_0$ and $o_1$. Concurrently, the right column subunits load $w^2_{0/1}$, $w^2_{1/1}$, $w^3_{0/1}$, and $w^3_{1/1}$, the MSBs of the weights from filters 2 and 3 and multiply them with $a_{0/0}$ and $a_{1/0}$. In \textbf{cycle 5} and Figure~\ref{fig:bitSerial_FC_example_f}, the right subunits complete the multiplication of their WR-held weights and $a_{0/1}$ and $a_{1/1}$ the MSBs of the two activations. By the end of this cycle, output activations $o_2$ and $o_3$ are ready as well.

In total it took 4+1 cycles to process 32 $1b \times 1b$ products (4, 8, 8, 8, 4 products in cycles 1 through 5, respectively). Notice that at the end of the 5th cycle, the left column subunits are idle, thus the WRs could have loaded another set of weights commencing the computation of a new set of outputs. In the steady state, with $2b$ input activations and weights, this engine will be producing  8 $1b \times 1b$ terms every cycle thus matching the 2 $2b \times 2b$ throughput of the parallel engine.  If the weights could be represented using only one bit, \BSD would be producing two output activations per cycle, twice the bandwidth of the bit-parallel engine. 

In general, if the bit-parallel hardware was using $P_{base}$ bits to represent the weights while only $P_{w}$ bits were actually required, for the \FCLs the \BSD engine would outperform the bit-parallel engine by $\frac{P_{base}}{P_{w}}$. The \BSD would use an array of $P_{base}\times k$ units, where $k$ the number of $P_{base}\times P_{base}$ products \BASE processes per cycle. Each subunit would produce $k$ $1b\times 1b$ products.  Since there is no weight reuse in \FCLs, $16$ cycles are required to load a different set of weights to each of the $16$ columns. Thus having activations that use less than $16$ bits would not improve performance (but could improve energy efficiency).  

\begin{figure*}
\centering
\subfloat[Baseline design ]{
\centering
\includegraphics[width=0.375\textwidth]{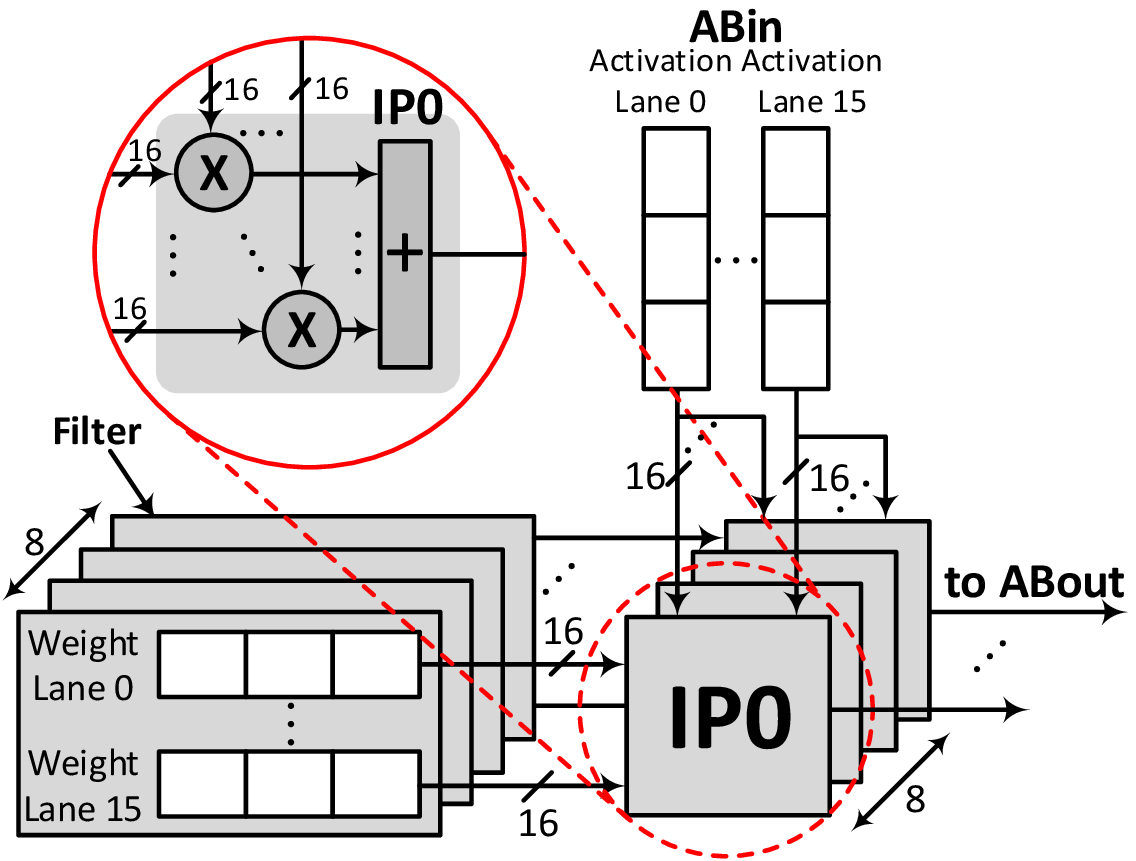}
\label{fig:DaDN-overview}
}
\subfloat[\BSDL ]{
\centering
\includegraphics[width=0.525\textwidth]{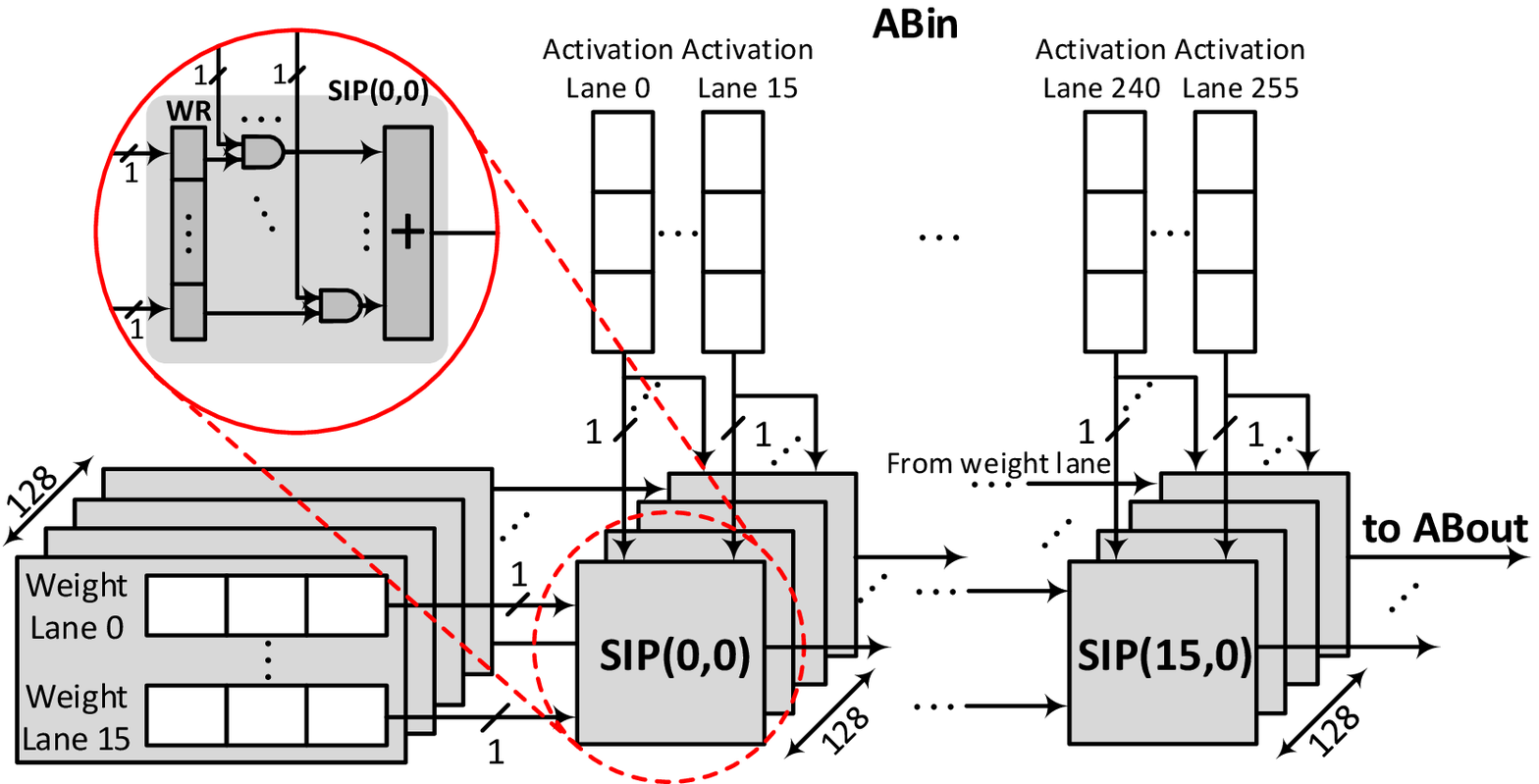}
\label{fig:BSD-overview}
}
\caption{The two CNN accelerators.} 
\end{figure*}

\noindent\textbf{Convolutional Layers:}
\BSD  processes \CVLs similarly to \FCLs but exploits weight reuse across different windows to exploit a reduction in precision for both weights and activations. Specifically, in \CVLs  the subunits across the same row share the same weight bits which they load in parallel into their WRs in a single cycle. These weight bits are multiplied by the corresponding activation bits over $P_a$ cycles. Another set of weight bits needs to be loaded every $P_{a}$ cycles, where $P_{a}$ is the input activation precision. Here \BSD exploits weight reuse across multiple windows by having each subunit column process a different set of activations. Assuming that the bit-parallel engine uses $P$ bits to represent both input activations and weights, \BSD will outperform the bit-parallel engine by $\frac{P^2}{P_{w} \times P_{a}}$ where $P_{w}$ and $P_{a}$ are the weight and activation precisions \BSD uses respectively.         

\section{Loom Architecture}
\label{sec:bsdArch}
This section describes the baseline fixed precision bit-parallel accelerator and the \BSDL architecture.

\subsection{Data Supply and Baseline System}
\label{sec:baseline}
Our baseline design (\BASE) shown on Figure~\ref{fig:DaDN-overview} is an appropriately configured data-parallel engine inspired by the DaDianNao accelerator~\cite{DaDiannao} the \textit{de facto} standard used for comparison in most accelerator studies. \BASE uses 16-bit fixed-point activations and weights. \BASE comprises $k$ inner product units (IP) each processing a different filter. Every cycle \BASE accepts as input $N$ activations and $N$ corresponding weights per filter out of $k$ filters. In the configuration shown $N=16$ and $k=8$. The $N$ activations are broadcast to all IP units. Each IP unit multiplies each of the $N$ activations with one out of its $N$ weights, reduces the resulting $N$ 32b products with an adder tree, and  accumulates the result into an output register. In total, every cycle, \BASE calculates $N\times k$ products producing $k$ partial output activations. 

An Activation Memory (AM) and a Weight Memory (WM) supply respectively the activations and the weights. An input activation buffer (\textit{ABin}) buffers the input activations while an output activation buffer (\textit{ABout}) temporarily buffers the output activations. For clarity, in our description we assume a single tile that processes up to 128 weights (8 filters) and 16 activations per cycle.  

\subsection{Loom}
For \BSD to match our \BASE configuration it needs to process 128 filters concurrently and 16 weight bits per filter per cycle, for a total of $128 \times 16 = 2048$ weight bits per cycle. Alternatively, \BSD could process 32 filters over 64 windows, however, we leave this investigation for future work. \BSD also accepts 256 1-bit input activations each of which it multiplies with 128 1-bit weights thus matching the computation bandwidth of base in the worst case where both activations and weights need 16 bits.   
Figure~\ref{fig:BSD-overview} shows the \BSDL design. It comprises 2K Serial Inner-Product Units (SIPs) organized in a $128\times 16$ grid. Every cycle, each SIP multiplies 16 $1b$ input activations with 16 $1b$ weights and reduces these products into a partial output activation. The SIPs along the same row share a common $16b$ weight bus, and the SIPs along the same column share a common $16b$ activation bus. Accordingly, as in \BASE, the SIP array is fed by a $2Kb$ weight bus and a $256b$ activation input bus. Similar to \BASE, \BSD has an ABout and an ABin. \BSD processes both activations and weights bit-serially.

\sloppy
\noindent\textbf{Reducing Memory Footprint and Bandwidth:} Since both weights and activations are processed bit-serially, \BSD can store weights and activations in a bit-interleaved fashion and using only as many bits as necessary thus boosting the effective bandwidth and storage capacity of the  weight memory and the AM. For example, given 2K $13b$ weights to be processed in parallel, \BSD would pack first their bit 0 onto continuous rows, then their bit 1, and so on up to bit 12. \BASE would stored them using 16 bits instead. A transposer can rotate the output activations prior to writing them to AM from ABout. Since each output activation entails inner-products with tens to hundreds of inputs, the transposer demand will be low. 

\noindent\textbf{Convolutional Layers:}
Processing starts by reading in parallel 2K weight bits from memory, loading 16 bits to all WRs per SIP row. The loaded weights will be multiplied by 16 corresponding activation bits per SIP column bit-serially over $P_{a}^L$ cycles where $P_{a}^L$ is the activation precision for this layer $L$. Then, the second bit of weights will be loaded into WRs and multiplied with another set of 16 activation bits per SIP row, and so on. In total, the bit-serial multiplication will take $P_{a}^L \times P_{w}^L$ cycles. where $P_w^L$ the weight precision for this layer $L$.  Whereas \BASE would process 16 sets of 16 activations and 128 filters over 256 cycles, \BSD processes them concurrently but bit-serially over $P_{a}^L \times P_{w}^L$ cycles. If $P_{a}^L$ and/or $P_{w}^L$ are less than 16, \BSD will outperform \BASE by $256/(P_{a}^L\times P_{w}^L)$. Otherwise, \BSD will match \BASE's performance. 

\noindent\textbf{Fully-Connected Layers:} Processing starts by loading the LSBs of a set of weights into the WR registers of the first SIP column and multiplying the loaded weights by the LSBs of the corresponding activations. In the second cycle, while the first column of SIPs is still busy with multiplying the LSBs of its WRs by the second bit of the activations, the LSBs of a new set of weights can be loaded into the WRs of the second SIP column. Each weight bit is reused for 16 cycles multiplying with bits 0 through bit 15 of the input activations. Thus, there is enough time for \BSD to keep any single column of SIPs busy while loading new sets of weights to the other 15 columns. For example, as shown in Figure~\ref{fig:BSD-overview} \BSD can load a single bit of 2K weights to SIP(0,0)..SIP(0,127) in cycle 0, then load a single-bit of the next 2K weights to SIP(1,0)..SIP(1,127) in cycle 1, and so on. After the first 15 cycles, all SIPs are fully utilized. It will take $P_{w}^L \times 16$ cycles for \BSD to process 16 sets of 16 activations and 128 filters while \BASE processes them in 256 cycles. Thus, when $P_{w}^L$ is less than 16, \BSD will outperform \BASE by $16/P_{w}^L$ and it will match \BASE's performance otherwise.

\begin{figure*}[!t]
\centering
\includegraphics[scale=0.8]{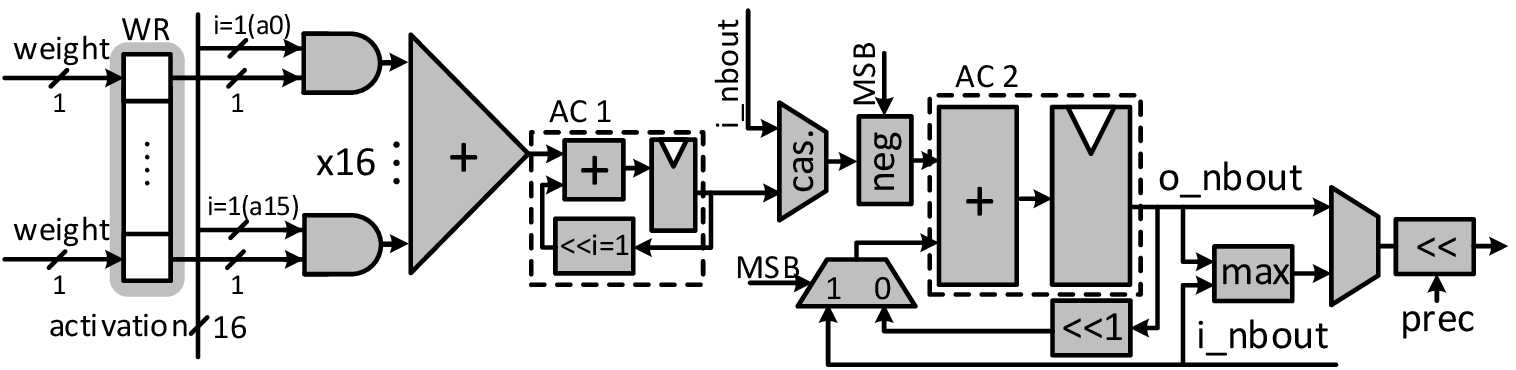}
\caption{\BSD's SIP.}
\label{fig:BSD-sip}
\end{figure*}

\begin{table}
    \centering
    \small
    \caption{
Activation and weight (W) precision profiles in bits for the convolutional and fully-connected layers. 
}
	\begin{tabular}{|l|p{2.4cm}|c|p{2.28cm}|c|} 
\hline
                   & \multicolumn{4}{c|}{\textbf{Convolutional Layers}}                                                                                                                                                           \\ 
\cline{2-2}\cline{3-3}\cline{4-4}\cline{5-5}
 \textbf{Network}  & \multicolumn{2}{c|}{\textbf{100\% Accuracy} }                                                          & \multicolumn{2}{c|}{\textbf{99\% Accuracy} }                                                        \\ 
\cline{2-2}\cline{3-5}
                   & {\textbf{Act. / Per Layer} }                         & {\textbf{W}} & {\textbf{Act. / Per Layer} }                     & {\textbf{W}}  \\ 
\hline
NiN                & {8-8-8-9-7-8-8-9-9-8-8-8}                         & {11}         & {8-8-7-9-7-8-8-9-9-8-7-8}                     & {10}          \\ 
\hline
AlexNet            & {9-8-5-5-7}                                       & {11}         & {9-7-4-5-7}                                   & {11}          \\ 
\hline
Google          & {10-8-10-9-8-10-9-8-9-10-7}                       & {11}         & {10-8-9-8-8-9-10-8-9-10-8}                    & {10}          \\ 
\hline
VGGS               & {7-8-9-7-9}                                       & {12}         & {7-8-9-7-9}                                   & {11}          \\ 
\hline
VGGM               & {7-7-7-8-7}                                       & {12}         & {6-8-7-7-7}                                   & {12}          \\ 
\hline
VGG19              & 12-12-12-11-12-10-11-11-\allowbreak13-12-13-13-13-13-13-13 & {12}         & {9-9-9-8-12-10-10-12-13-\allowbreak11-12-13-13-13-13-13} & {12}          \\ 
\hline
\hline
                   & \multicolumn{4}{c|}{\textbf{Fully-Connected Layers}}                                                                                                                                                         \\ 
\hline
                   & \textbf{100\% Accuracy}                                              &                                 & \textbf{99\% Accuracy}                                           &                                  \\ \hline

                   & \textbf{Weights /Per Layer}                                                &                                 & \textbf{Weights/Per Layer}                                       &                                  \\ 
\hline
NiN                & N/A                                                                  &                                 & N/A                                                              &                                  \\ 
\hline
AlexNet            & 10-9-9                                                               &                                 & 9-8-8                                                            &                                  \\ 
\hline
Google          & 7                                                                    &                                 & 7                                                                &                                  \\ 
\hline
VGGS               & 10-9-9                                                               &                                 & 9-9-8                                                            &                                  \\ 
\hline
VGGM               & 10-8-8                                                               &                                 & 9-8-8                                                            &                                  \\ 
\hline
VGG19              & 10-9-9                                                               &                                 & 10-9-8                                                           &                                  \\
\hline
\end{tabular}
\label{tab:CONV_precisions}
\end{table}

\noindent\textbf{SIP: Bit-Serial Inner-Product Units:}
\label{sec:SIP}
Figure~\ref{fig:BSD-sip} shows \BSD's Bit-Serial Inner-Product Unit (SIP). Every clock cycle, each SIP multiplies 16 single-bit activations by 16 single-bit weights to produce a partial output activation. Internally, each SIP has 16 1-bit Weight Registers (WRs), 16 2-input AND gates to multiply the weights in the WRs with the incoming input activation bits, and a 16-input $1b$ adder tree that sums these partial products. 
$AC_1$ accumulates and shifts the output of the adder tree over $P_{a}^L$ cycles. Every $P_{a}^L$ cycles, $AC_2$ shifts the output of $AC_1$ and accumulates it into the OR. After $P_{a}^L \times P_{w}^L$ cycles the Output Register (OR) contains the inner-product of an activation and weight set. In each SIP, a multiplexer after $AC_{1}$ implements cascading. To support signed 2's complement activations, a negation block is used to subtract the sum of the input activations corresponding to the most significant bit of weights (MSB) from the partial sum when the MSB is 1. Each SIP also includes a comparator (max) to support max pooling layers. 

\noindent\textbf{Dynamic Precision Reduction:}
\label{sec:Dynamic-Precision}
So far we assumed that software provided profile-derived per layer activation and weight precisions~\cite{judd:reduced}. Lascorz \textit{et al.,} observed that the hardware can further shorten these precisions by inspecting the actual values at runtime~\cite{dynamicstripes}. \BSD  determines adjusts precision per group of 256 activations that it processes concurrently. Per bit position OR trees produce a 16-bit vector indicating the positions where any of the activations has a 1. A leading one detector identifies the most significant position and thus the precision in bits that is sufficient.

\noindent\textbf{Processing Layers with Few Outputs: }For \BSD to keep all the SIPs busy an output activation must be assigned to each SIP. This is possible as long as the layer has at least 2K outputs. However, in the networks studied some \FCLs have only 1K output activations,  To avoid underutilization, \BSD's implements \textit{SIP cascading}, in which SIPs along each row can form a daisy-chain, where the output of one can feed into an input of the next via a multiplexer. This way, the computation of an output activation can be sliced along the bit dimension over the SIPs in the same row. In this case, each SIP processes only a portion of the input activations resulting into several partial output activations along the SIPs on the same row. Over the next $Sn$ cycles, where $Sn$ is the number of bit slices used, the $Sn$ partial outputs can be reduced into the final output activation.

\noindent\textbf{Other Layers:}
Similar to \DaDN, \BSD processes the additional layers needed by the studied networks. To do so, \BSD incorporates units for MAX pooling as in \DaDN. Moreover, to apply nonlinear activations, an activation functional unit is present at the output of the ABout. Given that each output activation typically takes several cycles to compute, it is not necessary to use more such functional units compared to \BASE.

\noindent\textbf{Total computational bandwidth: }
In the worst case, with 16b activations and weights, a single $16b \times 16b$ product that would have taken \BASE one cycle to produce, now takes LM 256 cycles. Since \BASE calculates 128 products per cycle, LM needs to calculate the equivalent of $256 \times 128$ $16b \times 16b$ products every 256 cycles. LM has $128 \times 16 = 2048$ SIPs each producing 16 $1b \times 1b$ products per cycle. Thus, over 256 cycles, LM produces $2048\times 16\times 256$ $1b \times 1b$ products matching \BASE's compute bandwidth. 

\noindent\textbf{Tuning the Performance, Area and Energy Trade-off:}
We can trade off some of the performance benefits to reduce the number of SIPs and the respective area overhead by processing multiple activation bits per cycle. 
The evaluation section considers 2-bit ($\BSD_{2b}$) and 4-bit ($\BSD_{4b}$) \BSD configurations which need 8 and 4 SIP columns and accommodate precisions that are multiple of 2 and 4, respectively. For example, for $\BSD_{4b}$ reducing the $P_a^L$ from 8 to 5 bits produces no performance benefit, whereas for the $\BSD_{1b}$ it would improve performance by $1.6\times$.

\section{Evaluation}
\label{sec:evaluation}
This section evaluates \BSDL performance, energy and area and explores the trade-off between accuracy and performance comparing to \BASE and \STR~\cite{Stripes-MICRO}.


\subsection{Methodology}
Execution time is modeled via a custom cycle-accurate simulator and energy and area measurements are collected over layouts of all designs. The designs were synthesized for worst case, typical case, and best case corners with the Synopsys Design Compiler using a TSMC 65nm library. Layouts were produced with Cadence Innovus using the typical corner case synthesis results which were more pessimistic for \BSD than the worst case scenario. Power results are based on the actual data-driven activity factors. The clock frequency of all designs is set to 1GHz. The ABin and ABout SRAM buffers were modeled with CACTI~\cite{Muralimanohar_cacti6.0:} and AM and WM were modeled as eDRAM  with Destiny~\cite{destiny}. We first evaluate \BSD assuming that all the activations fit on chip and the weights can be read from off-chip memory without any bandwidth constraint to explore the design space without being affected by the choice of a particular off-chip memory. We conclude by investigating performance with a single-channel of low-power DDR4-4267. 

\subsection{Weight and Activation Precisions: }
\label{sec:method-numerical}
Table~\ref{tab:CONV_precisions} reports the profile-derived per layer precisions of input activations and network precisions of weights for the \CVLs and \FCLs using the method of Judd \textit{et al.}~\cite{judd:reduced}. Since \BSD's performance for the \CVLs depends on both $P_{a}^L$ and $P_{w}^L$, we adjust them independently. We use per layer activation precisions and a common across all \CVLs weight precision. We found little inter-layer variability for weight precisions but additional per layer exploration is warranted. Since \BSD's performance for \FCLs performance depends only on $P_{w}^L$ we only adjust weight precision for \FCLs. The precisions that guarantee no \textit{top-1} accuracy loss for \CVLs input activations vary from 5 to 13 bits and for weights vary from 10 to 12. When a 99\% relative \textit{top-1} accuracy is still acceptable, the activation and weight precision can be as low as 4 and 10 bits, respectively. The per layer weight precisions for the \FCLs vary from 7 to 10 bits.

\input{tableh}


\begin{figure*}
\centering
\subfloat[\BSD's performance relative to \BASE.]{
\centering
\includegraphics[width=0.45\textwidth]{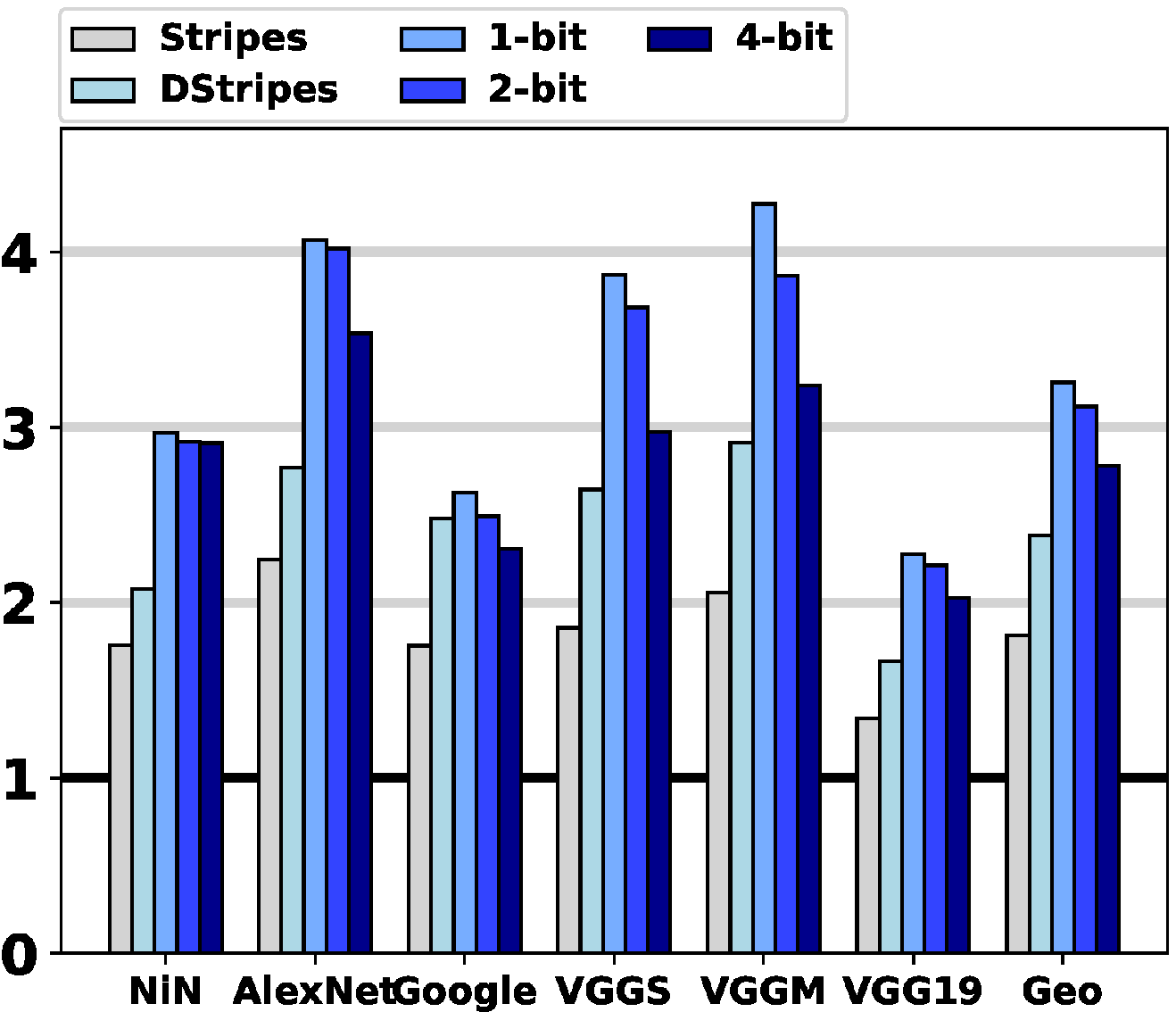}
\label{fig:all-speedup-dyn-100}
}
\subfloat[\BSD's energy efficiency relative to \BASE.]{
\centering
\includegraphics[width=0.47\textwidth]{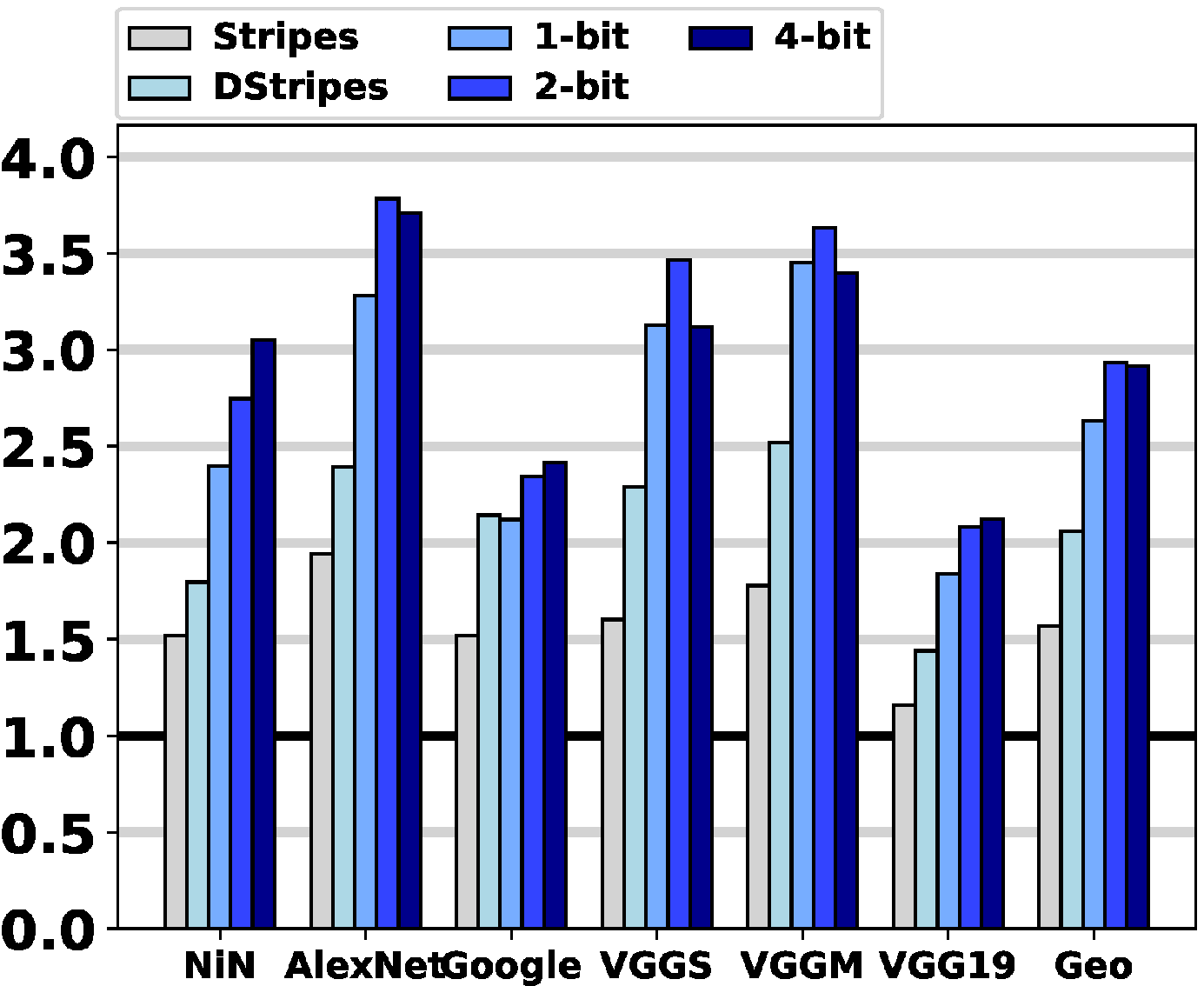}
\label{fig:all-eff-dyn-100}
}
\caption{\BSD's performance and energy efficiency relative to \BASE for all layers with 100\% accuracy.} 
\end{figure*}

\subsection{Performance and Energy Efficiency} 
Figures~\ref{fig:all-speedup-dyn-100} and~\ref{fig:all-eff-dyn-100} show respectively the performance and energy efficiency of \BSDL, \STR, and \STRD configurations relative to \BASE with the precision 100\% profiles of Table~\ref{tab:CONV_precisions} and for all layers combined. \STR is based on Stripes which exploits only profile-derived per layer activation precisions and only for CVLs~\cite{Stripes-MICRO}. \STRD incorporates dynamic prediction reduction~\cite{dynamicstripes}.

On average, ${\BSD}_{1b}$ outperforms \BASE by more than $3\times$ while being more than $2.5\times$ energy efficient. When \BSD processes multiple bits per cycle the performance benefits are lower but energy efficiency improves up to $2.9\times$. ${\BSD}_{1b}$ consistently outperforms \STR and \STRD in performance and \STR in energy efficiency. ${\BSD}_{1b}$ is more energy efficient than \STRD except for GoogleNet where its energy efficiency is within $2\%$ of \STRD.

Table~\ref{tbl:Perf:EE} reports per network performance and energy efficiency for \BSD configurations relative to \BASE for the \FCLs and \CVLs separately, and for the 100\% and 99\% accuracy profiles. In general, $\BSD_{1b}$ outperforms $\BSD_{2b}$ and $\BSD_{4b}$ in most cases with the latter two being more energy efficient. On occasion the latter two outperform $\BSD_{1b}$ under the 100\% accuracy profiles in \FCLs. 
Since for \BSD the performance improvement in \FCLs is only due to the use of lower weight precisions, processing multiple activation bits per cycle does not effect performance in the steady state. 
However, processing more activation bits per cycle reduces the initiation interval per layer an effect that becomes noticeable for small \FCLs. 

The table reports detailed results for \STR. For FCLs, \STR performance and energy efficiency suffer as it does not exploit weight precisions. With the 99\% accuracy profiles, both performance and energy efficiency improve considerably for FCLs and CVLs. Performance with \STRD would be identical to \STR for the \FCLs. We do not present detailed results for \STRD due to space limitations noting that \BSD consistently outperforms \STRD while being more energy efficient except for the \CVLs for GoogLeNet where the difference in energy efficiency is small.

\fixme{
\begin{table}
\centering
\caption{Average Effective Per Layer Weight Precisions~\cite{DPRed}}
\label{tbl:effprew}
\begin{tabular}{|l|l|}
\hline
\textbf{Network} & \textbf{Effective Precision Per Layer}                                                  \\ \hline
\textbf{NiN} & 8.85-10.29-10.21-7.65-9.13-9.04-7.63-\\&8.65-8.62-7.79-7.96-8.18                                                                \\ \hline
\textbf{AlexNet} & 8.36-7.62-7.62-7.44-7.55
\\ \hline 
\textbf{Google} & 6.19-5.75-6.80-6.28-5.34-6.70-6.31-5.02\\&-5.49-7.89-4.83 \\ \hline
\textbf{VGGS}  & 9.94-6.96-8.53-8.13-8.10
\\ \hline 
\textbf{VGGM}  & 9.87-7.55-8.52-8.16-8.14
\\ \hline 
\textbf{VGG19} & 10.98-9.81-9.31-9.09-8.58-8.04-7.89-7.86\\& -7.51-7.20-7.36-7.47-7.61-7.66-7.66-7.63
\\ \hline  
\end{tabular}
\end{table}
}

\subsection{Area Overhead} 
Post layout measurements were used to measure the area of \BASE and \BSDL. The $LM_{1b}$ configuration requires $1.34\times$ more area over \BASE while achieving on average a {\meanspeedupALLone} speedup. The $LM_{2b}$ and $LM_{4b}$ reduce the area overhead to $1.25\times$ and $1.16\times$ while still improving the execution time by {\meanspeedupALL} and {\meanspeedupALLfour}, respectively. Thus \BSD exhibits better performance vs. area scaling than \BASE.



 
\subsection{Scaling}
Thus far we assumed that all activations fit on chip and focused on a single \BSD configuration. We next consider configurations with practical on- and off-ship memory hierarchies.
Specifically, we size the activation memory so that most layers can fit on-chip avoiding off-chip accesses that today require at least two orders of magnitude more energy a critical consideration in embedded systems. Accordingly, \BASE requires 2MB of activation memory (VGG19 requires 10MB which is impractical for embedded systems and thus has to spill activations off-chip). Since \BSD processes both activations and weights bit-serially, it naturally stores and communicates values on- and off-chip using the per layer precisions. As a result, \BSD requires only 1MB on-chip memory for the activations. However, since \BSD processes more filters concurrently compared to \BASE, it can benefit from a larger weight memory. 

\begin{figure}[!t]
\centering
\includegraphics[scale=0.35]{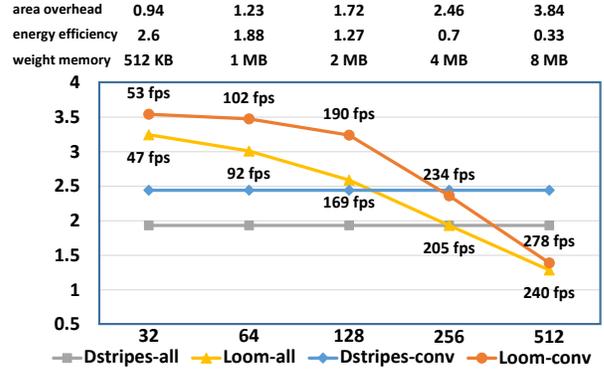}
\caption{Scaling vs. equivalent \BASE peak compute bandwidth. Conv: convolutional layers only. All: all layers. All results with a LPDDR4-4267 off-chip memory.}
\label{fig:perf:bw}
\end{figure}

Figure~\ref{fig:perf:bw} shows how average performance over all networks scales for different configurations where the number of SIPs is chosen to match the peak compute bandwidth (x-axis) of a bit-parallel accelerator. For example, the "128" configurations can perform the equivalent of 128 $16b\times16b$ multiply-accumulate operations per cycle. For each configuration Figure~\ref{fig:perf:bw} reports performance relative to \BASE and absolute performance as frames per second (fps). The figure reports results for the convolutional layers only and also for all layers. This is done because fully-connected layers are off-chip bound (and thus are affected by our choice of off-chip memory) whereas the convolutional layers are compute bound. Here we restrict attention to $\BSD_{1b}$. 

\BSD outperforms \BASE for all design points shown and can achieve real-time processing rates even for the "32" configuration. The relative performance advantage of \BSD drops for the larger configurations since \BSD requires more parallelism and suffers more from increased underutilization as the number of weight lanes grows. \STRD's relative performance over \BASE remains constant for the range shown. \BSD outperforms \STRD up to the "128" configurations. At "256" \BSD and \STRD perform nearly identically and at "512" the latter performs better. 

The figure also reports the weight memory capacity, the relative (vs. \BASE) area overhead, and the energy efficiency for the various \BSD configurations. For the "64" and "32" configurations \BSD requires $128KB$ and $544KB$ less memory in total than \BASE. However, for the "128" and the "256" configurations \BSD requires more memory than \BASE. Regardless, the performance benefits exceed the relative area overhead and thus \BSD provides a better performance/area trade-off than \BASE. For the "256" configuration energy efficiency suffers with \BSD. However, this measurement ignores the energy of off-chip traffic which is on average $0.61\times$ less with \BSD. Moreover, as CNNs evolve to process higher resolution images the size of activation memory increases significantly compared to the filter sizes which makes the effect of data compression more important~\cite{SSD}. Thus we expect that for higher resolution images \BSD will ever more appealing.

\begin{table}[ht!]
\centering
\small
\caption{Relative execution time speedup and energy efficiency with \BSD for all layers vs. \BASE.}
\label{tbl:effectiveWRes}
\begin{tabular}{|l|l|l|l|l|l|l|}
\hline
& \multicolumn{6}{|c|}{\textbf{All LAYERS COMBINED}}  \\ \cline{2-7}
\multirow{2}{*}{\textbf{Network}} & \multicolumn{2}{c|}{\textbf{\BSDL 1-bit}}  & \multicolumn{2}{c|}{\textbf{\BSDL 2-bit}}  & \multicolumn{2}{c|}{\textbf{\BSDL 4-bit}} \\ 
\cline{2-7} 
                         & \textbf{Perf}  & \textbf{Eff}  & \textbf{Perf} & \textbf{Eff}  & \textbf{Perf} & \textbf{Eff}  \\ 
 \cline{2-7}                        
& \multicolumn{6}{|c|}{\textbf{100\% TOP-1 Accuracy}}                                                                                      \\ \hline
NiN     & 3.38 & 2.73 & 3.32 & 3.13 & 3.31 & 3.48 \\ \hline
AlexNet & 5.66 & 4.57 & 5.61 & 4.57 & 4.95 & 5.19 \\ \hline
Google  & 3.19 & 2.57 & 3.02 & 2.84 & 2.80 & 2.93 \\ \hline
VGGS    & 5.72 & 4.62 & 5.46 & 5.13 & 4.42 & 4.63 \\ \hline
VGGM    & 6.03 & 4.87 & 5.46 & 5.14 & 4.60 & 4.83 \\ \hline
VGG19   & 3.38 & 2.73 & 3.28 & 3.09 & 3.01 & 3.15 \\ \hline
\textbf{Geomean} & \textbf{4.38} & \textbf{3.54} & \textbf{4.20} & \textbf{3.95} & \textbf{3.76} & \textbf{3.94} \\ \hline
\end{tabular}
\end{table}

%


\subsection{Per Group Weight Precisions}
Thus far we assumed that \BSD exploits software provided profile-derived per layer weight precisions~\cite{judd:reduced}. However, exploiting the approach of Lascorz et al.~\cite{DPRed} \BSD can further trim the weight precisions at a finer granularity to boost the performance and energy efficiency of both \FCLs and \CVLs. The per group weight precisions can be detected at runtime similarly to the activation precisions, or can be detected statically and communicated via per group metadata.

Table~\ref{tbl:effprew} reports the average effective weight precision per layer for a group of 16 weights. The \textit{estimated} performance and energy efficiency of \BSDL configurations relative to \BASE with the precision profiles of Table~\ref{tbl:effprew} and for all layers combined is shown in Table~\ref{tbl:effectiveWRes}. For these estimates we assume that performance scales linearly with weight precision. 

Exploiting the effective weight precisions yields a speedup of $4.38\times$, $4.20\times$, and $3.76\times$ over \BASE for $\BSD_{1b}$, $\BSD_{2b}$, and $\BSD_{4b}$ configurations, respectively. The energy efficiency of \BSD over \BASE is $3.54\times$, $3.95\times$, and $3.94\times$ for the aforementioned configurations.

\section{Related Work}
\label{sec:related} 
Due to space limitations, we limit attention to a few works that are the most related. We have already compared to \textit{Stripes}~\cite{Stripes-MICRO} extended with dynamic prediction reduction~\cite{dynamicstripes}.

\textit{Pragmatic}'s performance for the \CVLs depends only on the number of activation bits that are 1, but does not improve performance for \FCLs~\cite{pragmatic}. Further performance improvement may be possible by combining \textit{Pragmatic}'s approach with \BSD's but the costs per SIP may make this prohibitively expensive. \textit{Proteus} exploits per layer precisions reducing memory footprint and bandwidth but requires crossbars per input weight~\cite{judd2016proteus}. \BSDL does not need crossbars. Hardwired NN implementations  naturally exploit per layer precisions \cite{szabo_full-parallel_2000}. \BSDL does not require that the whole network fit on chip nor does it hardwire precisions. Furthermore, \BSDL further trims activations precisions at runtime.

Several accelerators target ineffectual weights and/or activations for dense and/or sparse networks~\cite{han_eie:isca_2016,albericio:cnvlutin,CambriconX,SCNN}. Most target either FCLs or CVLs alone. \BSD targets both layer types and benefits \textit{all} inputs ineffectual or not.  

\section{Conclusion}
\label{sec:conclusion}

This work presented \BSDL, a hardware inference accelerator for DNNs whose execution time for the convolutional and the fully-connected layers scales inversely proportionally with the precision $p$ used to represent the input data. \BSD can trade-off accuracy vs. performance and energy efficiency on the fly. 
Future work may consider extending \BSD to further exploit weight sparsity.

\begin{small}
\bibliographystyle{ieeetr}
\bibliography{sample-bibliography} 
\end{small}

%% file: tableh.tex
\begin{table*}[ht!]
\centering
\small
\caption{Relative execution time speedup and energy efficiency with \STR and \BSD for fully-connected and convolutional layers vs. \BASE.}
\label{tbl:Perf:EE}
\begin{tabular}{|l|l|l||l|l|l|l|l|l|l|l||l|l|l|l|l|l|}
\hline
& \multicolumn{8}{|c|}{\textbf{FULLY-CONNECTED LAYERS}}  & \multicolumn{8}{c|}{\textbf{CONVOLUTIONAL LAYERS}}                                                                                       \\ \cline{2-17}
\multirow{2}{*}{\textbf{Network}} & \multicolumn{2}{c||}{\textbf{Stripes}} & \multicolumn{2}{c|}{\textbf{\BSDL 1-bit}}  & \multicolumn{2}{c|}{\textbf{\BSDL 2-bit}}  & \multicolumn{2}{c|}{\textbf{\BSDL 4-bit}} & \multicolumn{2}{c||}{\textbf{Stripes}} & \multicolumn{2}{c|}{\textbf{\BSDL 1-bit}}  & \multicolumn{2}{c|}{\textbf{\BSDL 2-bit}}  & \multicolumn{2}{c|}{\textbf{\BSDL 4-bit}}                            \\ 
\cline{2-17} 
                         & \textbf{Perf}  & \textbf{Eff}  & \textbf{Perf} & \textbf{Eff}  & \textbf{Perf} & \textbf{Eff}  & \textbf{Perf} & \textbf{Eff} & \textbf{Perf}  & \textbf{Eff}  & \textbf{Perf} & \textbf{Eff}  & \textbf{Perf} & \textbf{Eff}  & \textbf{Perf} & \textbf{Eff} \\ 
 \cline{2-17}                        
& \multicolumn{8}{|c|}{\textbf{100\% TOP-1 Accuracy}}  & \multicolumn{8}{|c|}{\textbf{100\% TOP-1 Accuracy}}                                                                                     \\ \hline
NiN     & n/a  & n/a  & n/a  & n/a  & n/a  & n/a  & n/a  & n/a & 1.76 & 1.54 & 2.97 & 2.40 & 2.92 & 2.75 & 2.91 & 3.05 \\ \hline
AlexNet & 1.00 & 0.88 & 1.65 & 1.34 & 1.66 & 1.56 & 1.66 & 1.74 & 2.34 & 2.04 & 4.25 & 3.43 & 4.20 & 3.96 & 3.66 & 3.84 \\ \hline
Google  & 0.99 & 0.87 & 2.25 & 1.82 & 2.27 & 2.14 & 2.28 & 2.39 & 1.76 & 1.50 & 2.63 & 2.12 & 2.49 & 2.34 & 2.12 & 2.22 \\ \hline
VGGS    & 1.00 & 0.88 & 1.63 & 1.32 & 1.63 & 1.54 & 1.63 & 1.71 & 1.89 & 1.65 & 3.98 & 3.21 & 3.78 & 3.56 & 3.02 & 3.17 \\ \hline
VGGM    & 1.00 & 0.88 & 1.63 & 1.32 & 1.64 & 1.54 & 1.64 & 1.72 & 2.12 & 1.86 & 4.12 & 3.33 & 3.69 & 3.47 & 3.34 & 3.50 \\ \hline
VGG19   & 1.00 & 0.88 & 1.62 & 1.31 & 1.63 & 1.53 & 1.63 & 1.71 & 1.34 & 1.17 & 2.17 & 1.76 & 2.09 & 1.97 & 2.03 & 2.13 \\ \hline
\textbf{Geomean} & \textbf{1.00} & \textbf{0.88} & \textbf{1.74} & \textbf{1.41} & \textbf{1.75} & \textbf{1.65} & \textbf{1.75} & \textbf{1.84} & \textbf{1.84} & \textbf{1.61} & \textbf{3.25} & \textbf{2.63} & \textbf{3.10} & \textbf{2.92} & \textbf{2.78} & \textbf{2.92} \\ \hline
& \multicolumn{8}{|c|}{\textbf{99\% TOP-1 Accuracy}} & \multicolumn{8}{|c|}{\textbf{99\% TOP-1 Accuracy}}                                                                                      \\ \hline
NiN     & n/a  & n/a  & n/a  & n/a  & n/a  & n/a  & n/a  & n/a & 2.31 & 2.02 & 4.21 & 3.40 & 4.09 & 3.85 & 3.78 & 3.96 \\ \hline
AlexNet & 1.00 & 0.88 & 1.85 & 1.49 & 1.85 & 1.74 & 1.85 & 1.94 & 2.57 & 2.25 & 4.62 & 3.73 & 4.49 & 4.23 & 4.36 & 4.57 \\ \hline
Google  & 0.99 & 0.87 & 2.25 & 1.82 & 2.27 & 2.14 & 2.28 & 2.39 & 1.80 & 1.58 & 2.91 & 2.35 & 2.74 & 2.58 & 2.30 & 2.42 \\ \hline
VGGS    & 1.00 & 0.88 & 1.78 & 1.44 & 1.78 & 1.68 & 1.79 & 1.87 & 1.89 & 1.65 & 3.98 & 3.21 & 3.78 & 3.56 & 3.15 & 3.30 \\ \hline
VGGM    & 1.00 & 0.88 & 1.79 & 1.45 & 1.80 & 1.69 & 1.80 & 1.89 & 2.12 & 1.86 & 4.49 & 3.63 & 4.03 & 3.79 & 3.64 & 3.82 \\ \hline
VGG19   & 1.00 & 0.88 & 1.63 & 1.32 & 1.63 & 1.54 & 1.63 & 1.71 & 1.45 & 1.27 & 2.28 & 1.84 & 2.21 & 2.08 & 2.07 & 2.17 \\ \hline
\textbf{Geomean} & \textbf{1.00} & \textbf{0.88} & \textbf{1.85}  & \textbf{1.49} & \textbf{1.85} & \textbf{1.75} & \textbf{1.86} & \textbf{1.95} & \textbf{1.99} & \textbf{1.74} & \textbf{3.63} & \textbf{2.93} & \textbf{3.45} & \textbf{3.25} & \textbf{3.11} & \textbf{3.26} \\ \hline

\end{tabular}
\end{table*}